\documentclass[12pt,a4paper]{article}
\usepackage{amsbsy}
\usepackage{amsfonts}
\usepackage{cite}

\newcommand{\al}{\alpha}

\newcommand{\de}{\delta}

\newcommand{\cd}{\cdot}

\newcommand{\pa}{\partial}
\newcommand{\tfrac}[2]{{\textstyle\frac{#1}{#2}}}
\newcommand{\ee}{\mathrm{e}}
\newcommand{\ket}[1]{\vert #1 \rangle}
\newcommand{\bra}[1]{\langle #1 \vert}
\newcommand{\tr}{\mathrm{tr}\,}
\newcommand{\xb}{{\boldsymbol x}}

\newcommand{\pb}{{\boldsymbol p}}
\newcommand{\qb}{{\boldsymbol q}}
\newcommand{\rb}{{\boldsymbol r}}
\newcommand{\kb}{{\boldsymbol k}}

\newcommand{\Pp}{{\cal P}_{+}}
\newcommand{\Pm}{{\cal P}_{-}}

\newcommand{\ET}{\widetilde{E}}
\newcommand{\KT}{\widetilde{K}}
\newcommand{\VT}{\widetilde{V}}
\newcommand{\chiT}{\widetilde{\chi}}
\newcommand{\psiT}{\widetilde{\psi}}
\newcommand{\psiTp}{\widetilde{\psi}_{{+}}}
\newcommand{\psiTm}{\widetilde{\psi}_{{-}}}
\newcommand{\pslash}{p\!\!\!/}

\newcommand{\intdp}{\int\!\frac{\mathrm{d}^dp}{(2\pi)^d}\;}

\newcommand{\intdq}{\int\!\frac{\mathrm{d}^dq}{(2\pi)^d}\;}

\begin{document}
\begin{titlepage}

\begin{center}{\Large{\textbf{The Structure of Screening in QED}}}\\ [12truemm]
\textsc{Emili Bagan}\footnote{email: bagan@ifae.es}\\ [5truemm]
\textit{Dept.~Fisica Te\`orica \&\ IFAE\\
Edifici Cn\\
Universitat Aut\`onoma de Barcelona\\E-08193 Bellaterra  (Barcelona)\\
Spain}\\ [10truemm]\textsc{Martin Lavelle}\footnote{email:
mlavelle@plymouth.ac.uk} and \textsc{David
McMullan}\footnote{email: dmcmullan@plymouth.ac.uk}\\ [5truemm]
\textit{Department of
Mathematics and Statistics\\ The University of Plymouth\\
Plymouth, PL4 8AA\\ UK}\\ [10truemm] \textsc{Shogo
Tanimura}\footnote{email: tanimura@kues.kyoto-u.ac.jp}\\ [5truemm]
\textit{Dept. of Engineering Physics and Mechanics\\
Kyoto University\\
Kyoto 606-8501\\
Japan}
\end{center}

\bigskip
\begin{quote}
\textbf{Abstract:} The possibility of constructing charged
particles in gauge theories has long been the subject of debate.
In the context of QED we have shown how to construct operators
which have a particle description. In this paper we further
support this programme by showing how the screening interactions
arise between these charges. Unexpectedly we see that there are
two different gauge invariant contributions with opposite signs.
Their difference gives the expected result.
\end{quote}

\end{titlepage}

\subsubsection*{Introduction}

The long range nature of the electromagnetic interaction means
that the QED coupling cannot be naively switched off. Neglecting
this leads to the infra-red problem and the lack of a pole
structure in the on-shell Green's functions and $S$-matrix. This
has been taken \cite{kulish:1970ut} to mean that one cannot
describe charged particles in gauge theories. In a series of
papers
\cite{Lavelle:1997ty,Horan:1998im,Bagan:1999jf,Bagan:1999jk,Bagan:2000mk}
we have shown that this conclusion is overly hasty: it is in fact
possible to construct gauge invariant operators whose $S$-matrix
elements are free of infra-red divergences. These fields have been
shown to asymptotically recover a particle description of charges
and to have a rich structure which is physically reflected in the
cancellation of both soft and phase divergences. In QCD we have
demonstrated \cite{Bagan:2000nc,Lavelle:1998dv} (in both $2+1$
 and $3+1$ dimensions) that the term responsible for the cancellation
of soft divergences generates the anti-screening forces between
static quarks. We have then suggested that the factor responsible
for cancelling phase divergences must generate the screening
interaction. Here we will show that this is indeed the case.

\subsubsection*{The Structure of Static Charges}

Many years ago Dirac \cite{Dirac:1955uv} suggested that a static
charged particle should be described by the locally gauge
invariant operator
\begin{equation}\label{dirac}
 \psi_{\mathrm{D}}(x)\equiv \exp\left( -ie\frac{\partial_i
 A_i}{\nabla^2}(x)\right)\psi(x)
\,.
\end{equation}
His argument for this was that, in addition to the essential
requirement of gauge invariance, it has the expected equal-time
commutator with the electric field operator
\begin{equation}\label{elec}
 [E_i(x),\psi_{\mathrm{D}}(y)]
=-\frac {e}{4\pi} \frac{{x}_i-{y}_i}{\vert{\boldsymbol{x}-
\boldsymbol{y}}\vert^3}\psi_{\mathrm{D}}(y)\,,
\end{equation}
\textit{i.e.,} it recovers the static Coulombic electric field in
$3+1$ dimensions. This  argument also works in $2+1$ dimensions.

In \cite{Horan:1998im} we have shown that this electric field
requirement is not unique even at lowest order in the coupling. In
fact, arguing from a general kinematical point of view (inspired
by the heavy quark effective theory), we have shown that the
correct description of a \emph{static} charge is given by the
dressed field
\begin{eqnarray}\label{full}
  h^{-1}(x)\psi(x)&=& \ee^{-ieK(x)}\ee^{-ie\chi(x)}\psi(x)\nonumber
  \\[1mm]
  &=&
  \exp\bigg(\!\!-ie\int_{-\infty}^{x_0} \frac{\pa_i E_{i}}{\nabla^2}(s,\xb)\,ds\bigg)
  \exp\bigg(\!\!-ie\frac{\pa_iA_i}{\nabla^2}(x)\bigg)\psi(x)\,.
\end{eqnarray}
The new factor is separately gauge invariant and has a vanishing
commutator with the electric field in the absence of light
charges. We have shown that it (and its generalisation for a
moving charge) is essential in the cancellation of the phase
divergences associated with pair production processes.

In order to show that this provides the correct dynamical
description of physical charges, we have investigated the
potential between them. In the non-abelian theory we have
demonstrated \cite{Lavelle:1998dv,Bagan:2000nc} that the
generalisation to QCD of Dirac's proposal, \textit{i.e.,} just the
minimal part of the dressing,  produces the anti-screening
interaction at order $g^4$. We will now show that the new factor
in (\ref{full}) produces the screening effects at the same order
of perturbation theory.

\subsubsection*{The Potential Between Charges}

As usual we identify
\cite{Lavelle:1997ty,Lavelle:1998dv,Bagan:2000nc}  the potential
with the separation dependent part of the matrix element
\begin{equation}
  \bra0 h(y')h^{-1}(y)\,H_0\,h(y)h^{-1}(y')\ket0\,.
\end{equation}
For the purposes of this paper we can neglect higher terms in the
expansion of the dressing and simply write
\begin{equation}
  h^{-1}(y)=1-ie(K(y)+\chi(y))\,.
\end{equation}
Following  our discussion above, we will refer to the $K$ term as
the \emph{phase} contribution and $\chi$ as the \emph{soft}
structure.

The relevant part of the free Hamiltonian in $d$ spatial
dimensions is
\begin{eqnarray}
  H_0&=&\tfrac12\intdp\ET_i(\pb,x_0)\ET_i(-\pb,x_0)\\
       &&+\intdp E_p\left(
  \psiTm^\dag(-\pb,x_0)\psiTp(\pb,x_0)+\psiTm(\pb,x_0)\psiTp(-\pb,x_0)\right),
\end{eqnarray}
where we have dropped the irrelevant magnetic part of the
Hamiltonian and the terms involving the static charges. In the
second term here only light fermions are included and our positive
and negative frequency decomposition is defined by
\begin{eqnarray}
\psiT(\pb,x_0)&=&\frac1{\sqrt{2E_p}}\left(b^\al(\pb)u^\al(\pb)\ee^{-iE_px_0
}+{d^\al}^\dag(-\pb)v^\al(-\pb)\ee^{iE_px_0}\right)\nonumber\\
&=&\psiTp(\pb,x_0)+\psiTm(\pb,x_0)\,.
\end{eqnarray}
We recall that Gauss' law in momentum space reads:
\begin{equation}\label{gl}
  p_i\ET_i(\pb,x_0)=ie\intdq
  \psiT^\dag(\qb,x_0)\psiT(\pb-\qb,x_0)\,,
\end{equation}
where again, we neglect the heavy, static charges.
\subsubsection*{Lowest Order}

It is easy to see that at lowest order the momentum space
contribution comes from the commutator of the Hamiltonian with the
soft terms in the dressings:
\begin{eqnarray}\label{ss}
  \VT(\qb,\kb)&=&e^2\intdp\bra0[\ET_i(\pb,x_0),\chiT(\qb,x_0)][\ET_i(-\pb,x_0),\chiT(\kb,x_0)]
  \ket0\nonumber \\
  &=&-(2\pi)^de^2\de(\qb+\kb)\frac1{\qb^2}\,.
\end{eqnarray}
Performing the $\kb$ integral recovers the usual result
\begin{equation}\label{sss}
  \VT(\qb)=-e^2\frac1{\qb^2}=-\frac{4\pi\alpha}{\qb^2}\,.
\end{equation}
Note that this gives the correct $d$-dimensional
configuration space Coulombic potential between heavy charges at a separation $\rb$
\begin{equation}\label{born}
  V(\rb)=-e^2\frac{\Gamma(\frac{d}2-1)}{4\pi^{\frac{d}2}}
  \frac1{|\rb|^{d-2}}\,.
\end{equation}
The extension of this soft-soft contribution to the non-abelian
theory gives anti-screening.

In the absence of light charges (\ref{born}) is the full result in
QED and it is easy to see that there is no contribution from the
phase dressing. The presence of light fermions, however, modifies
the potential and we expect a screening effect. We now want to
show that our full dressing generates this screening force. There
are two contributions to this and we will analyse them in turn.

\subsubsection*{Phase-Phase Contribution}

The first term we want to calculate comes from the phase-phase
analogue of the soft-soft structure in (\ref{ss})
\begin{eqnarray}
&&\VT_{\mathrm{pp}}(\qb,\kb)=2e^2\intdp E_p\Biggl\{\Biggr.\tr
\bra0[\psiTm^\dag(-\pb,x_0),\KT(\qb,x_0)][\psiTp(\pb,x_0),
\KT(\kb,x_0)]\ket0\nonumber\\
&&\qquad\qquad\qquad\qquad+ \tr
\bra0[\psiTm(\pb,x_0),\KT(\qb,x_0)][\psiTp^\dag(-\pb,x_0),
\KT(\kb,x_0)]\ket0\Biggl.\Biggr\}\,.
\end{eqnarray}
After using Gauss' law (\ref{gl}) to rewrite the phase dressing as
\begin{equation}
  \KT(\pb,x_0)=-e\intdq\int_{-\infty}^{x_0}\mathrm{d}s
  \frac1{\pb^2}\psiT^\dag(\qb,s)\psiT(\pb-\qb,s)\,,
\end{equation}
we get
\begin{eqnarray}
\VT_{\mathrm{pp}}(\qb,\kb)&=&-2(2\pi)^de^4\frac1{\qb^4}\de(\qb+\kb)\intdp
\frac{E_p}{(E_p+E_{q-p})^2}\nonumber\\
&&\qquad\qquad\times\tr\Bigl(\Pm(\qb-\pb)\Pp(\pb)+\Pp(\qb-\pb)\Pm(\pb)\Bigr)\,,
\end{eqnarray}
where ${\cal P}_\pm(\pb)=(\pslash\pm m)\gamma^0/2E_p$ are the
projectors onto positive/negative frequencies.

 From the result that
\begin{equation}
  \tr\Bigl(\Pm(\pb)\Pp(\qb)\Bigr)=\frac{(d+1)n_{\mathrm{f}}}{2E_p2E_q}(E_pE_q+\pb\cd\qb-m^2)\,,
\end{equation}
where $n_{\mathrm{f}}$ is the number of light fermion species,
we can trivially integrate out $\kb$ to obtain the phase-phase
contribution to the potential at $e^4$
\begin{equation}
\VT_{\mathrm{pp}}(\qb)=-e^4(d+1)n_{\mathrm{f}}\frac1{\qb^4}\intdp
\frac{E_pE_{q-p}+\pb\cd(\qb-\pb)-m^2}{E_{q-p}(E_p+E_{q-p})^2}\,.
\end{equation}
Expanding around large $\pb$ here  gives the following divergent
correction in $d=3-2\epsilon$ dimensions
\begin{equation}\label{anti}
  \VT_{\mathrm{pp}}(\qb)=-\frac{4\pi\alpha_0}{\qb^2}\frac{\alpha_0}\pi
  \frac{n_{\mathrm{f}}}{3}\left[\frac1{\epsilon}-\ln\left(\frac{\qb^2}{\mu^2}\right)\right]\,.
\end{equation}
The sign here, however, corresponds to anti-screening!

\subsection*{Screening}

In addition to this phase-phase contribution, there are also two
identical soft-phase cross-terms. These structures yield
\begin{equation}
\VT_{\mathrm{sp}}(\qb,\kb)=2e^2\intdp\tr\bra0[\ET_i(\pb,x_0),
\chiT(\qb,x_0)][\ET_i(-\pb,x_0),\KT(\kb,x_0)]\ket0\,.
\end{equation}
Using Gauss' law (\ref{gl}) this becomes
\begin{eqnarray}
\VT_{\mathrm{sp}}(\qb,\kb)&=&2ie^4\frac1{\qb^2}\frac1{\kb^2}
\intdp\frac{\mathrm{d}^dp'}{(2\pi)^d}
\int_{-\infty}^{x_0}\mathrm{d}s\nonumber\\
&&\tr\bra0[\psiT^\dag(\pb,x_0)\psiT(\qb-\pb,x_0),
\psiT^\dag(\pb',s)\psiT(\kb-\pb',s)]\ket0\,.
\end{eqnarray}
After a little algebra, we obtain
\begin{eqnarray}
\VT_{\mathrm{sp}}(\qb,\kb)&=&2(2\pi)^de^4\frac1{\qb^4}
\de(\qb+\kb)\intdp\frac1{E_p+E_{q-p}}\nonumber\\
&&\qquad\times\tr\Bigl(\Pm(\pb)\Pp(\qb-\pb)+\Pm(-\pb)\Pp(\pb-\qb)\Bigr)\,.
\end{eqnarray}
This is then
\begin{equation}
\VT_{\mathrm{sp}}(\qb,\kb)=e^4(2\pi)^d(d+1)n_{\mathrm{f}}\frac1{\qb^4}\de(\qb+\kb)\intdp
\frac{E_pE_{q-p}+\pb\cd(\qb-\pb)-m^2}{E_pE_{q-p}(E_p+E_{q-p})}\,.
\end{equation}
Note that this is almost identical to the phase-phase term. The
only difference being the overall sign and the denominator term in
the momentum integral. In $d=3-2\epsilon$ dimensions, we find
\begin{equation}\label{scrn}
  \VT_{\mathrm{sp}}(\qb)=+\frac{4\pi\alpha_0}{\qb^2}\frac{2\alpha_0}\pi
  \frac{n_{\mathrm{f}}}{3}\left[\frac1{\epsilon}-\ln\left(\frac{\qb^2}{\mu^2}\right)\right]\,.
\end{equation}
Adding this to (\ref{anti}) we obtain the total (divergent)
contribution
\begin{equation}\label{tote}
  \VT(\qb)=-\frac{4\pi\alpha_0}{\qb^2}\left\{ 1-\frac{\alpha_0}
  \pi\frac{n_{\mathrm{f}}}{3}\left[\frac1{\epsilon}-\ln\left(\frac{\qb^2}{\mu^2}\right)\right]\right\}\,.
\end{equation}
 up to order $\alpha^2$. Charge renormalisation in QED corresponds to $\alpha_0=Z_\alpha\alpha$,
where
$Z_\alpha=1+\frac\alpha{\pi}\frac{n_{\mathrm{f}}}3\frac1\epsilon$.
We thus see that the divergences cancel as expected and we obtain
the usual screening result. We have thus  shown  at next to
leading order that the structures of the physical dressing
generate the interaction between charges.

\subsubsection*{Conclusion}

We have seen that the physical description of a  charge which we
propose indeed contains the effects which screen static charges.
In a concrete calculation we have seen that the overall screening
forces between such charges arise from two distinct, gauge
invariant contributions. One has an anti-screening effect, but it
is only half the size of the dominant screening term. This
separation is not apparent in other methods (such as Wilson
loops~\cite{Appelquist:1978es,Schroder:1998vy} and
non-relativistic perturbation
theory~\cite{Gribov:1977mi,Drell:1981gu}) and it is intriguing to
speculate on similar structures in the gluonic screening of QCD.

This result is a further vindication of our approach to the
fundamental question of how to describe charged particles in gauge
theories. We have seen that, from general principles, the dressing
around a charge has a rich structure which is reflected in the
infra-red properties of the fields and in the forces between
charges. This shows a previously unobserved intimate connection
between the soft structures ($\chi$) of gauge theories and
anti-screening and also between the phase structures ($K$) and the
overall screening effect.

\noindent\textbf{Acknowledgements:} This work was supported by the
British Council/Spanish Education Ministry \textit{Acciones
Integradas} grant no.\  1801/HB1997-0141, a Royal Society Joint
project grant with Japan and a PPARC Theory Travel Fund award.


\end{document}